\documentclass[conference]{IEEEtran}

\usepackage{subfiles}

\usepackage[utf8]{inputenc} 
\usepackage[T1]{fontenc}
\usepackage{url}
\usepackage{ifthen}
\usepackage{cite}
\usepackage[cmex10]{amsmath} 
\usepackage{xcolor}
\usepackage{tabularx,ragged2e,booktabs}
\usepackage{times}
\usepackage{float}
\usepackage{afterpage}
\usepackage{amstext}
\usepackage{amssymb,bm}
\usepackage{latexsym}
\usepackage{color}
\usepackage{pstricks}
\usepackage{enumerate}
\usepackage{mathtools}
\usepackage{epsfig}
\usepackage{afterpage}
\usepackage{amstext}
\usepackage{amssymb}
\usepackage{latexsym}
\usepackage{color}
\usepackage{graphicx}
\usepackage{amsthm}
\usepackage{graphicx}
\usepackage{pstricks}
\usepackage{booktabs}
\usepackage{enumerate}
\usepackage{subfig}
\usepackage{fixltx2e}
\usepackage{enumitem}
\usepackage{multirow}
\usepackage{verbatim}
\usepackage{mathabx}
\usepackage{hyperref}
\usepackage[ruled,vlined]{algorithm2e}

\DeclarePairedDelimiter\ceil{\lceil}{\rceil}
\DeclarePairedDelimiter\floor{\lfloor}{\rfloor}

\newtheorem{thm}{Theorem}

\newtheorem{prop}{Proposition}

\newcommand{\cah}{circular-arc hypergraph} 

\newcommand{\cardi}{t} 
\newcommand{\mbit}{\kappa} 
\newcommand{\Mod}[1]{\ (\mathrm{mod}\ #1)}
\newcommand{\remain}{r} 
\newcommand{\gnum}{\floor{\frac{m}{s}}} 
\newcommand{\spa}[1]{\text{Span}(#1)} 

\begin{document}
\title{Private Pliable Index Coding} 

\author{%
\IEEEauthorblockN{%
Tang Liu and Daniela Tuninetti\\%
University of Illinois at Chicago, Chicago, IL 60607 USA, \\
Email: {\tt tliu44, danielat@uic.edu}\\%
}%
}

\maketitle

\begin{abstract}
    The Pliable Index CODing (PICOD) problem is a variant of the Index Coding (IC) problem, where the desired messages by the users, who are equipped with message side information, is part of the optimization.
    This paper studies the PICOD problem where users are subject to a \emph{privacy constraint}. 
    In particular, the following spacial class of private PICODs is investigated:
    1) the side information structure is circular, and 
    2) each user can decode one and only one message.
    The first condition is a special case of the ``circular-arc network topology hypergraph'' class of PICOD studied in~\cite{consecutive_picod}, for which an optimal solution was given without the privacy constraint.
    The second condition was first studied in~\cite{secure_picod_achievability} and was motivated by the need to keep content privacy is some distribution networks.
    
    This paper proposes both converse and achievable bounds.
    The proposed achievable scheme not only strictly outperforms the one in~\cite{secure_picod_achievability} for some values of the system parameters, but it is also information theoretically optimal in some settings. For the remaining cases, the proposed linear code is shown to require at most one more transmission than the converse bound derived by restricting the sender to only use linear codes.
\end{abstract}

\section{Introduction}
\label{sec:introduction}

\paragraph{Pliable Index Coding (PICOD)}
\label{par:picod_introduction}
PICOD is a variant of the Index Coding (IC) problem and was first introduced in~\cite{BrahmaFragouli-IT1115-7254174}. In PICOD, the messages to be decoded by the users, who have message side information, are not part of the problem definition. Instead, in PICOD, the sender assigns to the users the messages they need to decode so that (i) the assigned messages were not already present in the local side information, and (ii) the length of the code that allows every user to recover the assigned message has the shortest possible length.
The PICOD problem formulation captures the nature of some content delivery applications, where there is flexibility in the choice of the desired messages to be delivered to the users. This flexibility allows to reduce the number of transmissions compared to an IC with the same side information structure. 

The IC problem in its general form is known to be hard~\cite{index_coding_with_sideinfo}.
The general PICOD problem is not simpler than the IC problem in terms of complexity. 
For instance, the linear PICOD (here the sender is restricted to use linear codes) is still NP-hard~\cite{Content_type_coding}.
Some efficient algorithms to solve the general PICOD were proposed  in~\cite{polytime_alg_picod}.
For the case where the side information structure of the PICOD has ``symmetry,'' we found the optimal code length (under no restriction of encoding scheme that the sender can use) in~\cite{consecutive_picod}.
However, the general PICOD problem is open. 

\paragraph{Private PICOD}
\label{par:past_works}

The problem of security and privacy in IC has been studied from different perspectives.
In~\cite{security_ic_with_sideinfo_eavesdropper}, the Authors proposed an IC model where an eavesdropper has a limited access to the side information sets and to the transmitted codeword; the goal here is to prevent the eavesdropper from obtaining any new information. 
In~\cite{private_bc_ic_approach}, the Authors considered an IC model  where the sender must design a code that allows each user to decode its desired message, but at the same time prevent him from obtaining any information about the side information or the desired messages of the other users.
This latter model has the flavor of the private information retrieval problem~\cite{private_information_retrieval}, where a user wants to hide its desired message and/or side information from the other users and the server.
Similarly to the private information retrieval problem, the Authors of~\cite{private_ic} formulated the \emph{private IC} problem, where a user in the IC problem should be able to decode only its own desired messages but no others.

Recently, in~\cite{secure_picod_achievability}, the Authors extended the private IC problem in~\cite{private_ic} to the PICOD framework. 
Only the case where the side information structure is ``circular'', and where each user can decode one and only one message was considered in~\cite{secure_picod_achievability}. 
Several schemes were given in~\cite{secure_picod_achievability} and shown to provide the desired level of privacy, but the optimality is discussed only under the linear encoding constraint for some cases.

\paragraph{Contributions and Paper Organization}
\label{par:contrib}

In this paper we study a generalization (in terms of the form of the side information sets) of the private PICOD model from~\cite{secure_picod_achievability}, as formally described in Section~\ref{sec:system_model}.
We provide both achievable and the converse bounds, where past work only focused on linear achievable schemes. 
The main result of this paper is presented and discussed in Section~\ref{sec:main_result}.
In Section~\ref{sec:proof} we derive both information theoretic and linear-code restricted converse bounds.
We also provide linear achievable schemes and show they are either information theoretically optimal, or differ from the linear-code restricted converse by at most one transmission.
Section~\ref{sec:conclusion} concludes the paper.
Some proofs are in Appendix.

\section{System Model}
\label{sec:system_model}

A private $(n,m,\mathcal{A})$ PICOD$(\cardi)$ is defined as follows.
There are $n\in\mathbb{N}$ users and one central transmitter. 
The user set is denoted as $U := \left\{ u_{1},u_{2},\ldots,u_{n} \right\}$.
There are $m\in\mathbb{N}$ independent and uniformly distributed binary messages of $\mbit \in \mathbb{N}$ bits each. The message set is denoted as $\mathcal{W} := \left\{ w_{1},w_{2},\ldots,w_{m} \right\}$.

The central transmitter has knowledge of all messages $\mathcal{W}$.
User $u_i$ has the messages indexed by its side information set $A_i\subset [m]$, $i\in[n]$. 
The messages index by $A_i$ are denoted as $W_{A_i}$. 
The collection of all side information sets is denoted as $\mathcal{A} := \{A_{1},A_{2},\ldots,A_{n}\}$, which is assumed globally known at all users and the transmitter. 

The sender and the users are connected by an error-free broadcast link. 
The sender transmits the codeword 
\begin{align}
    x^{\mbit\ell} :=  \mathsf{ENC} (\mathcal{W}, \mathcal{A}),
\end{align}
where $\mathsf{ENC}$ is the encoding function. 

The decoding function for user $u_j$ is 
\begin{align}
    \{\widehat{w}^{(j)}_{1},\dots,\widehat{w}^{(j)}_{\cardi}\} 
    := \mathsf{DEC}_j(W_{A_j},x^{ \mbit\ell}), \ \forall j\in [n],
\end{align}
where $\cardi$ is the number of messages desired by a user and not already included in $A_j$.
In other words, the decoding function at $u_j$ is $\mathsf{DEC}_j, \ j\in[n],$ such that
\begin{align}
	\label{eq:succeful_decoding}
    \Pr[ & \exists \{d_{j,1},\dots,d_{j,\cardi}\}\cap A_j=\emptyset :  \nonumber \\
    & \{\widehat{w}^{(j)}_{1},\dots,\widehat{w}^{(j)}_{\cardi}\}\neq \{{w}_{d_{j,1}},\dots,{w}_{d_{j,\cardi}}\}]  \leq \epsilon,
\end{align}
for some $\epsilon\in(0,1)$ and some $D_j := \{d_{j,1},\dots,d_{j,\cardi}\} \subseteq [m]\setminus{A_i}$. 
We set $D_j$ contains the indices of the desired messages by $u_j$. 

Up to this point, the system definition is that of a classical PICOD problem. We introduce now the privacy constraint. 
Privacy is modeled here as follows: user $u_j$ can not decode any messages other than the $\cardi$ messages indexed by $D_j$. 
Specifically, we impose that for all $j\in [n]$,
\begin{align}
	\label{eq:secure_constraint}
     &H( w_i  | x^{ \mbit\ell}, W_{A_j}, \mathcal{A} ) \nonumber \\
     \geq &  H( w_i) - \kappa \epsilon, \forall i\in [m]\setminus{(D_j \cup A_j)}.
\end{align}

A code is called \emph{valid} for the private $(n,m,\mathcal{A})$  PICOD$(\cardi)$ if and only if it satisfies the conditions in~\eqref{eq:succeful_decoding} and~\eqref{eq:secure_constraint}. The goal is to find a valid code and a desired message assignment that result in the smallest possible codelength, i.e., 
\begin{align}
     \ell^{\star}:= \min\{\ell : \text{$\exists$ a valid $x^{\mbit\ell}$ for some $\mbit$}\}. 
 \end{align}

Finally, if the encoding function at the sender is restricted to be a linear map from the message set, 
the length of shortest possible such valid codewords is denoted as 
$\ell^{\star}_{\mathrm{lin}}$.

\subsection{Network Topology Hypergraph (NTH) and size-$s$ circular-$h$ shift Side Information} 
\label{sub:circular_side_information}
In the rest of the paper we shall consider a class of $(n,m,\mathcal{A})$ private PICOD$(\cardi)$ problems with a specific structure on $\mathcal{A}$. Such class is a generalization of the one studied in the past work~\cite{secure_picod_achievability}, which is a special case of the circular-arc NTH that we studied in~\cite{consecutive_picod}, where we fully solved the case $\cardi=1$ for the circular-arc NTH without the privacy constraint.
The rest of the section contains graph definition that will be used later on.

Let $H=(V,\mathcal{E})$ denote a \emph{hypergraph} with vertex set $V$ and edge set $\mathcal{E}$, where an edge $E\in\mathcal{E}$ is a subset of $V$.
The NTH, first introduced in~\cite{consecutive_picod}, is a generalization of network topology graph for the IC problem~\cite{TIM_indexcoding}.
In a NTH, messages are the hyperedges, while the users are the vertices.
A user does \emph{NOT} have a message in its side information set if and only if its corresponding vertex is incident to the hyperedge that represents the message.
A $1$-\emph{factor} of $H$ is a spanning edge induced subgraph of $H$ that is $1$-regular.
A hypergraph $H$ is called an \emph{\cah} if there exists an ordering of the vertices $v_1, v_2, \ldots, v_n$ such that if $v_i, v_j, i\leq j$, then the $v_q$ for either all $i\leq q\leq j$, or all $q\leq i$ and $q\geq j$, are incident to an edge $E$.

In this paper we study the $(n,m,\mathcal{A})$ private  PICOD$(1)$ with a special side information set structure: the sets in $\mathcal{A}$ are size-$s$ circular-$h$ shift of the message set.
More precisely,
The side information set of user $u_i$ is of the form 
\begin{align}
A_i = \{ (i-1)h+1, \ldots, (i-1)h+s \}, 
\label{eq:size-s shift-h def}
\end{align}
for $i\in[n]$ where all indices are intended modulo the size of the message set, i.e., denoted as $\Mod m$ when needed,
where $0\leq s \leq m-\cardi$ and $h\geq 1$, here $\cardi=1$.

Let $g:=\gcd(m,h)$. In this private PICOD$(1)$ there are $n=m/g$ users, since all users have distinct side information sets.
Note that the size-$s$ circular-$h$ shift side information setup is a special case of the side information structure with \emph{circular-arc} we introduced in~\cite{consecutive_picod}.
Also, the model studied in~\cite{secure_picod_achievability} is the special case when $g=1$ (and thus $n=m$).

\section{Main Result} 
\label{sec:main_result}


For the size-$s$ circular-$h$ shift side information private PICOD$(1)$ problem, we have the following main result.
\begin{thm}
	\label{thm:constant_gap_1}
	For the private PICOD$(1)$ where the side information sets are 
	as in~\eqref{eq:size-s shift-h def} we have the following.
	
	Impossibility: when $m$ is odd, $g=1$, and either $s=m-2$ or $s=1$, a valid code does not exists  (i.e., it is not possible to satisfy the privacy constraint). 
	
	For the remaining possible cases, we have:
	    \begin{itemize}
	        \item For $s\geq m/2$, and either $1 \leq s< m/2, g \geq 3$, or $1\leq s< m/2, s\neq 2, g=2$
                	\begin{align}
            	    \label{eq:it_tight_case}
            	    \ell^* = 
                	    \begin{cases}
                	    1, & \text{\rm if the NTH has a $1$-factor,}\\
                	    2, & \text{\rm otherwise.}
                	    \end{cases}
            	    \end{align}
            \item For $1 \leq s<m/2$, and either $g =1$ or $s=g=2$
               	\begin{align}
                	\label{eq:linear_bounds}
                	    \ceil{\gnum/2}
                	    \leq \ell^{\star}_{\mathrm{lin}} 
                	    \leq 
                	    \begin{cases}
                	    \ceil{\gnum/2},   & \frac{m}{s} \in \mathbb{Z},\\
                	    \ceil{\gnum/2}+1, & \frac{m}{s} \notin \mathbb{Z}.
                	    \end{cases}
                	\end{align}
            \end{itemize}
\end{thm}

    A few observations are in order.
    When $s\geq m/2$, the achievable scheme provided in~\cite{secure_picod_achievability} is indeed information theoretical optimal given~\eqref{eq:it_tight_case}, which is our converse bound in~\cite[Theorem~3]{consecutive_picod} for the case without privacy constraint. 
    Therefore, our main contribution in Theorem~\ref{thm:constant_gap_1} is three-fold compared to~\cite{secure_picod_achievability}:
        1) for $s\geq m/2$ we provide information theoretic optimality of the scheme in~\cite{secure_picod_achievability}; 
        2) for $s<m/2$ we provide a new achievable scheme, and show it is almost linear optimal; 
        3) we generalize the side information structure to any $g>1$.

    In~\eqref{eq:linear_bounds}, 
    if we fix $s$ and $g$, $\gnum$ is monotonic in the message set size $m$.
    One interesting observation is that, although the lower bound on $\ell^{\star}_{\mathrm{lin}}$ is monotonic with $m$, the upper bound is not. 
    For instance, consider the case $s=2, g=1$; 
    when $m=10$ or $m=12$, we have $\ell^{\star}_{\mathrm{lin}}\leq 3$, while 
    when $m=11$ we have $\ell^{\star}_{\mathrm{lin}}\leq 4$.
    In other words, from the point of $m=11$, both increasing and decreasing the message set size may result in an increase of the required number of transmissions. 
    Note that this is the point where the upper and the lower bounds differ. 
    It is not clear at this point whether this means the achievable scheme here is not optimal, or the optimal private \emph{linear} PICOD solution is not monotonic in $m$. 



\section{Proof of Theorem~\ref{thm:constant_gap_1}} 
\label{sec:proof}

We divide the proof of Theorem~\ref{thm:constant_gap_1} into various cases.
Specifically, 
the impossibility result is proved in Section~\ref{sub:impossible_cases},
the case $s<m/2,g=1$ in Section~\ref{sub:case_s<m/2_g=1}, and
the case $s<m/2,g=s=2$ in Section~\ref{sub:brief_case_s<m/2_g=2_s=2}.
The schemes that achieve~\eqref{eq:it_tight_case} are sketched in Section~\ref{sub:brief_proof_remain_cases},
while the full proof can be found in Appendix~\ref{sec:appendix_proof_remain_cases}.

\subsection{Impossible Cases}
\label{sub:impossible_cases}

First we show that in some cases the privacy constraint can not be satisfied.
The proof of the same under a linear encoding constraint was provided in~\cite{secure_picod_achievability}.
Here we provide a simple information theoretic proof of the same.
The main idea is to proof the existence of a ``decoding chain'' (as defined in~\cite{consecutive_picod}) regardless of the choices of the desired messages at the users.
%
%
%
This ``decoding chain'' technique was used in~\cite{consecutive_picod} for the converse proof of so called consecutive complete--$S$ PICOD$(t)$. Since this argument does not rely on any assumption on the encoding function at the server, the resulting bound is truly information theoretical (as opposed to a form of `restricted converse').

    \subsubsection{Case $m$ is odd, $s=m-2$, and $g=1$}
    \label{ssub:impossibility_m_odd_s=m-2_g=1}
    User $u_i$ has two possible choices for its desired message (because all the others are in its side information set); these messages are $d_i=(i+s)\Mod m$ or $d_i=(i-1)\Mod m$.
    If $d_i=(i+s)\Mod m$, by decoding $w_{d_i}$, user $u_i$ can mimic $u_{(i-1)\Mod m}$ since $A_{(i-1)\Mod m}\subset \{(i+s)\Mod m\}\cup A_i$.
    Therefore, user $u_i$ can decode $w_{d_{(i-1)\Mod m}}$. 
    To make sure user $u_i$ can decode only one message, we need $d_{(i-1)\Mod m}\in A_i$ so that user $u_i$ does not decode another message that is not in its side information set.
    We thus have $d_i\in A_{(i-1)\Mod m}$ and $d_{(i-1)\Mod m}\in A_i$ can mimic each other.
    We say that two user mimicking each other form a ``loop''.
    The same argument holds for the other choice of $d_i$ as well.
    To make sure all users can decode one message only, every user must be in a ``loop''. 
    However, one user can be in only one loop.
    Thus, there must be one user that is not contained in any loop because here we have taken $m$ to be odd. 
    Therefore, there exists one user that can mimic another user and thus decode two messages, which violates the privacy constraint.

    \subsubsection{Case $m$ is odd, $s=1$, and $g=1$}
    \label{ssub:impossibility_m_odd_s=1_g=1}
    User $u_i$, by decoding its desired message $d_i=j, j\neq i$, can mimic user $u_j$ and thus also decode $d_j$. 
    To make sure user $u_i$ can decode only one message, we must have $d_j=i$.
    Therefore user $u_i$ and $u_j$ form a ``loop''.
    Similarly, every user can be in only one loop.
    We need all users to be in a loop to make sure that every user can decode at most one message.
    Since $m$ is odd, this is impossible. Thus, there must exists one user that can decode two messages, which violates the privacy constraint.

\subsection{Case $s<m/2$ and $g = 1$ (here $m=n$)}
\label{sub:case_s<m/2_g=1}

\subsubsection{Achievability}
\label{ssub:achievability_case_s<m/2_g=1}

	Let $m = 2sq+\remain$ for some $q,\remain\in \mathbb{Z}$ such that $0\leq \remain<2s$, i.e., $\remain$ is the remainder of $m$ modulo $2s$, and $q$ is the maximum number of users who can have disjoint side information sets. 
	We can have $2q+\floor{\frac{\remain}{s}}$ groups of $s$ users such that the users in each group have at least one message in common in their side information sets.
	Also, $\remain-s\floor{\frac{\remain}{s}}$ is the number of users that are not contained in any of these groups.

    The intuition of our achievable scheme is as follows.
    Under the privacy constraint, we can satisfy the users in two groups with one transmission, therefore $2sq$ users can be satisfied by $q$ transmissions. 
    If $\remain=0$, $q$ transmissions suffice;
    if $0<\remain\leq s$, we can satisfy the remaining $\remain$ users by one transmission; and
    if $s<\remain<2s$, we can satisfy the remaining $\remain$ users by two transmissions.
    Therefore the total number of transmissions is $q+\ceil{\frac{\remain}{s}}$.
	Based on this intuition, we distinguish three sub-cases: 
	a) $\remain=0$;
	b) $0<\remain\leq s$; and
	c) $s<\remain<2s$.

	\paragraph*{Case $\remain=0$} 
	\label{par:remainder=0}
	This is the case where $m$ is divisible by $2s$, therefore is divisible by $s$. 
	We partition the users into groups $G_1,G_2,\ldots,G_{2q}$, such that all users in $G_i$ have message $w_{is}$ in their side information.
	Set the desired message of the users in $G_{2i},i\in[q],$ to be $w_{(2i-1)s}$, and the desired message of the users in $G_{2i-1}, i\in[q]$ to be $w_{2is}$ 
	There are $q$ transmissions, each of them is $w_{2is}+w_{(2i-1)s}, i\in[q],$ that satisfies the users in $G_i$ and $G_{i+1}$ while
	it  does not provide any useful information for the users in other groups. 
	Therefore, $q=\frac{m}{2s}$ transmissions suffice to satisfy all the $m$ users.

	\paragraph*{Case $0<\remain \leq s$} 
	\label{par:0<remainder<=s}
	We partition the users into $2q+1$ groups.
	As for to the case $\remain=0$, the first $2q$ groups contain $s$ users. 
	The users in $G_i, i\in[2q],$ all have $w_{is}$ in their side information.
	Group $G_{2q+1}$ has $\remain$ users. 
	The first $q$ transmissions are $w_{2is}+w_{(2i-1)s}, i\in [q]$, and satisfy the users in groups $G_i,i\in[2q]$.
	We next satisfy the users in $G_{2q+1}$.
	
	If $\remain=1$, we have $G_{2q+1}=\{u_m\}$.
	Let $d_m=s+1$ and the $(q+1)$-th transmission be $w_{s+1}+\sum_{j\in A_m} w_j$.
	Note that 
	$s\geq \remain+1=2$, therefore user $u_m$ can decode $w_{s+1}$ while the other users can not decode any new messages one they receive the last transmission.
	
	If $\remain \geq 2$,
	the users in $G_{2q+1}$ all have $W_{[1:s-\remain]\cup\{m\}}$ in their side information. 
	Let $d_{2sq+1}=s-\remain+1$ and $d_j=2sq+1, j\in[2sq+2:m]$.
%
	The $(q+1)$-th transmission is $w_{2sq+1}+w_m+\sum_{j=1}^{s-\remain+1} w_{j}$.
	Since user $u_{2sq+1}$ can compute $w_{2sq+1}+w_m+\sum_{j=1}^{s-\remain} w_{j}$ and
	users $u_j,j\in[2sq+2:m],$ can compute $w_m+\sum_{j=1}^{s-\remain+1} w_{j}$, these users have the message that is not in their side information set as their desired message.
	All the other users who are not in $G_{2q+1}$ have at least two messages unknown in the transmission and thus cannot decode it.
	Therefore, each user can decode only one message by the achievable scheme with $q+1$ transmissions.
	If $m$ is divisible by $s$, then $\remain=s$ and $q+1=\ceil{\frac{m}{2s}}$;
	if $m$ is not divisible by $s$, $q+1=\ceil{\floor{\frac{m}{s}}/2}+1$. 
	
	\paragraph*{Case $s<\remain<2s$} 
	\label{par:s<remainder<2s}
	We partition the users into $2q+2$ groups. 
	The users in group $G_{i}, i\in[2q+1],$ all have message $w_{(is)}$, while the users in group $G_{2q+2}$ all have $ W_{[1: 2s-\remain]\cup\{m\}}$.
	We satisfy the first $2q$ groups 
	by sending $w_{2is}+w_{(2i-1)s}, i\in[q]$.
	We satisfy all users in $G_{2q+1}$ by sending $w_{2sq+1}+w_{2sq+s}+w_{2sq+s+1}$.
	%
	If $\remain=s+1$, $G_{2q+2}=\{u_m\}$ and we let $d_m=s+1$ and send as last transmission $w_{s+1}+\sum_{j\in A_m}$;
	otherwise, we let $d_{2sq+s+1}=2s-\remain+1$ and $d_j=2sq+s+1, j\in[2sq+s+1 : m]$ 
	and send $w_{2sq+s+1}+w_m+\sum_{i=1}^{2s-\remain+1} w_i$. 
	One can verify that all users can decode one and only one message by using a code of length $q+2=\ceil{\floor{\frac{m}{s}}/2}+1$.

\subsubsection{Converse}
\label{ssub:converse_case_s<m/2_g=1}

Messages are bit vectors of length $\mbit$, for some  $\mbit$;
we thus see each message as an element in $\mathbb{F}_{2^\mbit}$.
When the sender uses a linear code (on $\mathbb{F}_{2^\mbit}$), we can write the transmitted codeword as $x^{\ell} = E w^{ m}$, where $w^{m} = (w_1,w_2,\dots,w_m)^T$ is the vector containing all the messages, and where $E\in \mathbb{F}_{2^\mbit}^{\ell \times m}$ is the generator matrix of the code.
We denote the linear span of the row vectors of $E$ as $\spa{E}$. 
Recall that in this setting, user $u_i, i\in[n],$ must to be able to decode one and only one message outside its side information set $A_i$; the index of the decoded message is $d_i$.
%
Let $v_{i,j}$ be a vector whose $j$-th element is non-zero and all elements with index not in $A_i$ are zeros.

A valid generator matrix $E$ must satisfy the following two conditions:
\begin{enumerate}
    \item \emph{Decodability}: $v_{i,d_i}\in \spa{E}$, for all $i\in [m]$;
    \item \emph{Privacy}: $v_{i,j}\notin \spa{E}$ for all $i\in [m], j\in[m]\setminus (A_i\cup \{d_i\})$.
\end{enumerate}
The decodability condition guarantees successful decoding of the desired message $w_{d_i}$ by user $u_i$ as argued in~\cite{index_coding_with_sideinfo}.
The privacy condition must hold because
the existence of a vector $v_{i,j}\in \spa{E}$ for some $j\in[m]\setminus (A_i\cup \{d_i\})$ implies that user $u_i$ is able to decode message $w_j$ in addition to its desired message $w_{d_i}$.


The optimal linear code length $\ell^{\star}_{\mathrm{lin}}$ 
is the smallest rank of the generator matrix $E$, which by definition is the maximum number of pairwise linearly independent vectors in $\spa{E}$. We prove the linear converse bound by giving a lowered bound on the maximum number of pairwise linearly independent vectors in $\spa{E}$, i.e., the rank of $E$.
To do so, we need the following two propositions, proved in Appendices~\ref{app:p1} and~\ref{app:p2}, respectively. These propositions are the key technical novelty of this work.

\begin{prop}
\label{prop:trival_satisfaction_impossible}
In a working system (where every user can decode without violating the privacy condition) with $g=1$ we must have $e_i\notin\spa{E}$ for all $i\in [m]$, where $e_i$ are standard bases of $m$-dimensional linear space.
\end{prop}

\begin{prop}
\label{prop:linear_half_converse}
For a working system with $g=1$, among all $n$ users, consider $k$ users whose side information sets are pairwise disjoint. 
The number of transmissions of any linear code that satisfies these $k$ users must be $\ell_\mathrm{lin} \geq \ceil{k/2}$.
\end{prop}

Proposition~\ref{prop:trival_satisfaction_impossible} states that in this case,  a trivial `uncoded scheme' (that consists of sending $\ell^{\star}_{\mathrm{lin}}$  messages one by one) always violates the privacy constraint. 
In other words, no user is allowed to decode without using its side information.

Proposition~\ref{prop:linear_half_converse} provides a lower bound on the code-length of a linear code for a subset of the users in the system (those with pairwise disjoint side information sets), thus for all users.
Therefore, among all $m$ users in the system, there are $\floor{\frac{m}{s}}$ users with pairwise disjoint side information sets.
By Proposition~\ref{prop:linear_half_converse}, we need at least $\ceil{\floor{\frac{m}{s}}/2}$ transmissions to satisfy these users.
Therefore, in order to satisfy all the users in the system, we must have $\ell^{\star}_{\mathrm{lin}} \geq  \ceil{\floor{\frac{m}{s}}/2}$. 
This provides the claimed lower bound. 



\subsection{Case $s<m/2$ and $g=s=2$ (here $n=m/2$)}
\label{sub:brief_case_s<m/2_g=2_s=2}

\subsubsection{Achievability}
\label{ssub:achievability_case_s<m/2_g=2_s=2}
In this case we show $\ell^{\star}_{\mathrm{lin}} = \ceil{m/4}$.
We use the achievable scheme for case $s=2 < m/2$ and $g=1$ from Section~\ref{ssub:achievability_case_s<m/2_g=1}, where we need $\ceil{m/4}$ transmissions to satisfy all $n=m$ users. 
We users we have in this case are a proper subset of the users in the case $g=1$. 
The achievable scheme for $g=1$ still satisfies all users and meets the privacy constraint.
We have $\ell\leq \ceil{m/4}$ in this case.

\subsubsection{Converse}
\label{ssub:converse_case_s<m/2_g=2_s=2}
The converse proof in Section~\ref{ssub:converse_case_s<m/2_g=1} does not directly apply in this case, mainly because the proof of Proposition~\ref{prop:trival_satisfaction_impossible} requires $g=1$. In Appendix~\ref{app:p3} we show that it also holds for $g=2$, stated as Proposition~\ref{prop:trival_satisfaction_impossible_s=g=2}.

Hence the converse follows the same argument in Section~\ref{ssub:converse_case_s<m/2_g=1} by replacing Proposition~\ref{prop:trival_satisfaction_impossible} with Proposition~\ref{prop:trival_satisfaction_impossible_s=g=2}  in Appendix~\ref{app:p3}.
We show that for $k$ user with pairwise disjoint side information sets,  $\ceil{k/2}$ transmissions are needed for this case under the linear encoding restriction.
Note that in this case all $n=m/2$ users are with pairwise disjoint side information sets.
Therefore, the total number of transmissions that satisfy all users is at least $\ceil{m/4}$.


\subsection{Remaining Cases}
\label{sub:brief_proof_remain_cases}
We aim to prove~\eqref{eq:it_tight_case}.
Here we provide the converse proof, and a sketch of the achievability proofs. 
The detailed proofs can be found in Appendix~\ref{sec:appendix_proof_remain_cases}.

\subsubsection{Converse}
\label{ssub:remain_converse}
By the converse bound in~\cite[Theorem~3]{consecutive_picod} for the circular-arc PICOD($1$) without the privacy constraint, we have $\ell*\geq 1$ when the NTH has 1-factor, and $\ell^*\geq 2$ when the NTH has no 1-factor. This converse bound holds also when we impose an additional privacy constraint.

\subsubsection{Achievability for $s<m/2$, either $g=2,s\neq 2$, or $g \geq 3$}
\label{ssub:brief_case_s<m/2_g=2_s!=2_or_g>=3}
We show how to find the first message to transmit. Then, all the users that do not have this message in their side information sets must be satisfied by a second transmission. We show how to find this second transmission in such a way that the privacy constraint is met.

\subsubsection{Achievability for $s\geq m/2$}
\label{ssub:brief_case_s>=m/2}
The achievable scheme in this case is the one proposed in~\cite{secure_picod_achievability}, where only the case $g=1$ was considered.
For the cases where $g>1$, the set of users in the system is a proper subset of the set of users when $g=1$. 
Therefore the scheme for $g=1$ is still valid for any $g$ in that both decoding and privacy constraints are met.

\section{Conclusion}
\label{sec:conclusion}
In this paper we gave both achievable and converse bounds for the  private PICOD$(1)$ problem with circular side information sets. 
We showed that our linear achievable scheme is information theoretical optimal for some parameters, or it requires at most one more transmission compared to a converse developed under the constraint that the sender is restricted to use linear codes. 
Proving, or disproving, that our linear codes are actually information theoretically optimal is subject of current investigation.

This work was supported in part by NSF Award number 1527059. The opinion expressed in this paper are of the authors and do not necessarily reflect those of the NSF.

\appendices


\section{Proof of Proposition~\ref{prop:trival_satisfaction_impossible}}
\label{app:p1}
Recall that, for $g=1$, the side information sets are $A_i =(i,\ldots,i+s-1\Mod m)$ for all $i\in[m]$, as here $n=m$.
The proof is by contradiction. 
Assume without loss of generality (wlog) that we have a working systems with $e_1\in\spa{E}$, that is, every user can decode message $w_1$ without even using its side information. 
Then, all users $u_i, i\in [2:m-s+1]$ (who do not have $w_1$ in their side information sets) must have desired message $w_1$, in order to make sure that privacy constraint is not violated. 
This implies Fact~1: user $u_1$ can only have $w_{d_1}=w_{s+1}$ as desired message. 

Fact~1 is true because $u_2$ desires $w_1$, therefore $A_2 \cup \{d_2\} \supset A_1$. 
After decoding $w_1$, user $u_2$ can mimic user $u_1$ and thus decode message $d_2$. 
Since user $u_2$ can decode only one message, then $d_1\in A_2\setminus A_1 = \{s+1\}$. 
Therefore $d_1=s+1$.
By taking $d_1=s+1$, we conclude that there must exist vector $v_{1,d_1} =  v_{1,s+1} = c+\alpha_{s+1}e_{s+1}$,
where $\alpha \in \mathbb{F}_{2^\mbit}, \alpha\neq 0$ and $c\in \spa{A_{1}}$, where with an abuse of notation we let $\spa{A_i}$ denote $\spa{\{e_j : j\in A_i\}}$.

Given that we established Fact~1, let $j$ be the position of the fist non-zero element in the so found $v_{1,s+1}$.
Clearly, $j\leq s+1$ since the $(s+1)$-th element of $v_{1,s+1}$ is $\alpha_{s+1}\neq 0$. We have the following cases:
\begin{enumerate}
    \item If $j=s+1$, all the users who do not have $w_{s+1}$ in their side information sets, can decode $w_{s+1}$. 
    This is because in this case $ v_{1,s+1} = \alpha e_{s+1}$.
    Thus user $u_{s+2}$, who has neither $w_1$ nor $w_{s+1}$ in its side information set, can decode both $w_1$ and $w_{s+1}$.
    \item If $1<j<s+1$, then user $u_{j+1}$ can decode $w_j$, since $s+1\in A_j$. But user $u_{j+1}$ decodes $w_1$ by assumption.
    Therefore, user $u_j$ can decode both $w_1$ and $w_j$. 
    \item If $j=1$, user $u_{s+2}$ can decode both $w_{s+1}$ and $w_1$. 
    Therefore, $u_{s+2}$ can decode two messages. 
\end{enumerate}
In all the three above cases, there exists at least one user who can decode at least two messages, thus violating the privacy constraint. 
Therefore, the original assumption $e_1\in\spa{E}$ must be impossible in a working system. The same reasoning applies to any $e_j, j\in[m]$. This proves the claim.

\section{Proof of Proposition~\ref{prop:linear_half_converse}}
\label{app:p2}

By Proposition~\ref{prop:trival_satisfaction_impossible}, for all $i\in[k]$ there exists 
$v_{i,d_i} = \alpha_i e_{d_i}+c_i\in \spa{E}$, where $c_i\in \spa{A_i}$ and $\alpha_i \neq 0$.
Since the side information sets $A_i$ are assumed to be disjoint, the vectors $c_i$ are linearly independent.
$v_{i,d_i}$ are linearly dependent only if $d_i\in A_j$ and $d_j\in A_i$ for some $i\neq j$. 
In other words, there exists a ``loop'' between $u_i$ and $u_j$.
Note that since the side information sets are disjoint, one user can be in at most one ``loop'', and the number of ``loops'' is at most $\floor{k/2}$. 
Therefore the number of $v_{i,d_i}$ that are linearly dependent is at most $\floor{k/2}$, and thus the number of\emph{ linearly independent} $v_{i,d_i}$ is at least $k-\floor{k/2}=\ceil{k/2}$.
Therefore, the number of transmissions that is needed to satisfy $k$ users with disjoint side information sets must satisfy $\ell = \text{rk}(E) \geq \ceil{k/2}$.

\section{Proof of Proposition~\ref{prop:trival_satisfaction_impossible_s=g=2}}
\label{app:p3}
\begin{prop}
\label{prop:trival_satisfaction_impossible_s=g=2}
In a working system (where every user can decode without violating the privacy condition) with $g=s=2$ we must have $e_i\notin\spa{E}$ for all $i\in [m]$, where $e_i$ are standard bases of $m$-dimensional linear space.
\end{prop}

Similar to the proof of Proposition~\ref{prop:trival_satisfaction_impossible},
Wlog assume $e_1$ is in $\text{Span}(E)$. 
All users $u_i, i\in [2:m-s+1]$ in this case need to desire message $w_1$.
Let $d_1 \in A_j$, for some $j\neq 1$
For the decoding at $u_1$, there exists a vector $v_{1, d_1}\in \spa{E}$ such that: 
1) the $d_1$-th element is non-zero;
2) all elements with indices that are not $1,2$ or $d_1$ are zeros.
We check the first and second element of $v_{1, d_1}$ and have the following cases:
\begin{enumerate}
    \item Both the first and second elements of $v_{1, d_1}$ are zeros, $v_{1, d_1} = e_{d_i}$. 
    Therefore all users without $w_{d_i}$ in their side information sets can decode $w_{d_i}$. 
    \item The first element is zero while the second element is non-zero.
    By $v_{1, d_1}$ the user $u_j$ is able decode $w_2$ since $u_j$ already decodes $w_1$ and has $w_{d_1}$ in its side information sets. $u_j$ can decodes two messages.
    \item The first element is non-zero while the second element is zero.
    Since all users that do not have $w_1$ can decode $w_1$, all users can decode $w_{d_1}$ if they do not have it in their side information sets.
    \item Both the first and second elements of $v_{1, d_1}$ are non-zeros.
    $u_j$ decodes $w_1$ by assumption. It also has $w_{d_i}$ in its side information set. Therefore $u_j$ can decode $w_2$.
\end{enumerate}
All possible cases show that there exists at least one user that can decode at least two messages. 
The assumption that $e_1$ is in $\text{Span}(E)$ is impossible.
The reasoning applies to all $e_j,j\in [m]$. 
Therefore we conclude that $e_i\notin \spa{E}$ for all $i\in[m]$.

\section{Proof for the Remaining Cases}
\label{sec:appendix_proof_remain_cases}

For the following three cases: $s<m/2$, $g=2$, $s\neq 2$; $s<m/2$, $g \geq 3$; $s\geq m/2$, we aim to prove 
\begin{align*}
            	    \ell^* = 
                	    \begin{cases}
                	    1, & \text{if the NTH has $1$-factor,}\\
                	    2, & \text{otherwise.}
                	    \end{cases}
            	    \end{align*}

\subsection{Converse for all three cases}
\label{sub:remain_converse}
By the converse bound in ~\cite[Theorem~3]{consecutive_picod} for circular-arc PICOD, without the privacy constraint, $\ell*\geq 1$ when the NTH has 1-factor, and $\ell^*\geq 2$ when the NTH has no 1-factor. 

\subsection{Achievability for case $s<m/2$, $g = 2$, and $s\neq 2$}
\label{sub:remain_case_s<m/2_g=2_s!=2}

If $s=1$, the NTH has 1-factor. Thus $\ell^*=1$, in which case we send the sum of all messages.
If $2< s<m/2$, we send $w_{s+1}$ as the first transmission.
This transmission satisfies all users but $u_i, i = 2,\dots, \floor{s/2}+1$, since they all have $w_{s+1}$ in their side information set.
When $s$ is even, they have common side information set $\{s+1,s+2\}$.
We send the second transmission as $w_3+w_{s+1}+w_{s+2}+w_{s+3}$. $u_2$ can decode $w_{s+3}$, $u_i,i=3,\dots, \floor{s/2}+1$ can decode $w_3$. 
All the other users, after decoding $w_{s+1}$, still have at least two messages known in the summation, therefore can not decode any more messages.
When $s$ is odd, we send the second transmission as $w_3+w_{s}+w_{s+1}+w_{s+2}+w_{s+3}$.
By similar argument we can show that $u_i, i = 2,\dots, \floor{s/2}+1$ can decode one messages from the second transmission while the other users can not.


\subsection{Achievability for case $s<m/2$, $g\geq 3$}
\label{sub:remain_case_s<m/2_g>=3}

It is trivial that if the NTH has $1$-factor we have $\ell^*=1$, in which case we send the sum of all messages.
Therefore, we show that if the NTH does not have $1$-factor we can satisfy all users with two transmissions while satisfying the privacy constraint.
Send $w_{s+1}$ as the first transmission. 
All users who do not have $w_{s+1}$ in the side information sets are satisfied. 
The users that have $w_{s+1}$ in the side information sets are $u_i, i=2,\dots,\floor{s/g},\floor{s/g}+1$.
They have common side information set $[\floor{s/g}g+1 : s+g]$.
$|[s+2 : s+g]|\geq 2$ since $g\geq 3$.
For the second transmission we send $w_{m}+\sum_{i=s+2}^{s+g}w_{i}$.
By the condition $s<m/2$, all users $u_i, i=2,\dots,\floor{s/g},\floor{s/g}+1$ do not have $w_m$ in the side information sets.
Therefore these users can decode $w_m$ as the desired message.
For the second transmission, all the other users have at least two messages known in the summation, therefore can not decode any information from the second transmission.
The privacy constraint is satisfied.


\subsection{Achievability for case $s\geq m/2$}
\label{sub:remain_case_s>=m/2}


We use the proposed achievable scheme in \cite{secure_picod_achievability} for this case. 
When $g=1$, \cite{secure_picod_achievability} showed one can achieve $\ell=1$ if the NTH has 1-factor, and $\ell=2$ otherwise.
When $g>1$, the users are in a proper subset of the users of $g=1$.
Therefore the users can still be satisfied by the scheme that can satisfy strictly more users.
The privacy constraint is still satisfied as less users can not decode more messages.
Therefore, the achievable scheme can achieve $\ell=1$ when NTH has 1-factor, and $\ell=2$ when NTH does not have 1-factor.

\bibliographystyle{IEEEtranS}
\bibliography{refs}

\end{document}